# Paired Insulators and High Temperature Superconductors

By
T. H. Geballe and S. A. Kivelson
Stanford University

In common with all condensed matter physicists of our generations, our way of thinking about our field was shaped and greatly inspired by countless seminal works of Phil Anderson – a debt we are pleased to have the opportunity to acknowledge. Discussing plans for this article, we spent many pleasant times debating which particular contribution to highlight – super-exchange (THG), Anderson-Higgs (SAK), the Anderson-Morel pseudopotential (THG), poor man's scaling (SAK), etc. In the end, we opted to highlight a single specific paper which greatly affected each of us at the time, and which has continued to exert a strong intellectual influence on us in the ensuing years. Almost 40 years ago in Ref. [1], Phil introduced the negative U center to account for the fact that most glasses and amorphous semiconductors are diamagnetic. This paper has been highly influential, but certainly does not rank among Phil's most famous works; however, focusing on it enables us to reacquaint a younger generation with another of Phil's contributions, and to use this as a springboard to discuss some forward looking extensions that continue to fascinate us.

The inspiration for this work – as with much of Phil's work – was a set of simple experimental facts that make the conclusion almost self-evident when brought into conjunction by the master: Many amorphous semiconductors are highly insulating, even at room temperature, despite the fact that there is direct evidence of a large density of states at the Fermi energy; this indicates that the states at the Fermi energy must be strongly localized. None-the-less, these materials are often diamagnetic (exhibiting neither Curie nor Pauli paramagnetism), which implies that there must be a "spin-gap" of sorts. These observations can be reconciled, Phil observed, if there is a constant density of localized states with a strongly attractive negative U, so that in equilibrium, each state is either empty or occupied by a singlet pair of electrons. Assuming that the negative U must, in turn, be the consequence of a strong local electron-phonon coupling, he noted that if it is derived from a Holstein model, there is an accompanying exponentially large Franck-Condon reduction of the effective tunneling rate between neighboring localized states, which accounts for the absence of any measurable (hoping) conductance within the band of localized states.

The experiments Phil had in mind included measurements on amorphous and glassy semiconductors. In particular he noted that amorphous silicon is paramagnetic whereas chalcogenide glasses, such as germanium selenide, are diamagnetic. At an intuitive level, he suggested associating the localized states with dangling bonds which are occupied by single electrons in Si (where, presumably, U is positive) and by electron pairs in the chalcogenide glasses, where they are negative U centers.

In an aside in his paper, Phil observed that there is an analogy between the existence of an effective attraction between electrons in a superconductor and the negative U centers in an insulating glass. What is left unsaid is that the resulting spin-gap in the superconductor is measured in degrees, and sets the scale of the superconducting $T_c$, while in the glasses it is measured in electron volts and is related to the activation energy for conduction. In a conventional superconductor, increasing the strength of the electron-phonon coupling leads to an enhancement of $T_c$, while in the glass, it leads to an exponential increase of the Franck-Condon factor and hence increasingly strong localization. In the superconductor, the phonon frequency, $\omega_0$, is large compared to the spin-gap and $T_c$ is an increasing function of $\omega_0$, while in the glass, $\omega_0$ is small compared to the spin-gap and the Franck-Condon factor is a decreasing function of $\omega_0$. Still, it is hard not to fantasize that the high pairing scale in the glass could somehow be retained in a related conducting phase, where it would become the superconducting gap scale.

Following this idea and the discovery[2] of superconductivity in PbTe lightly doped with Tl, recent work [3] by one of us and collaborators has uncovered many reasons for believing that the superconductivity in Tl doped PbTe may be caused by negative U centers: 1) Doping PbTe with other cations results only in a doped semiconductor even at carrier concentrations at which superconductivity occurs on Tl doping. 2) The $T_c$ of $Pb_{1-x}Tl_xTe$ is more than an order of magnitude higher than in other low-density semiconductor-superconductors with similar carrier densities. 3) $T_c$ and a temperature dependence of the resistance consistent with scattering from a charge-Kondo impurity both onset above the same characteristic Tl concentration, $x_c \sim 0.03$. 4) The Hall number is proportional to x for $x < x_c$ and then becomes roughly x independent for larger x. Moreover, the same model has been successful in motivating and accounting for experiments [4] on the superconducting state of In doped SnTe.

In private correspondence with THG while this work was being carried out, Phil initially expressed interested skepticism. He particularly emphasized the fact that when the negative U results from lattice polarization, the Franck-Condon effect should drastically quench the "charge-Kondo" coupling, *i.e.* the coherent exchange of electron pairs between the localized centers and the conduction band. (His doubts were largely assuaged when further investigation found a minimum in the low temperature resistance of the sort expected from charge Kondo scattering.) There are two possible ways around this rather fundamental issue: 1) Very high frequency phonons could, in principle, mediate a strong attraction without an accompanying large Franck-Condon suppression of coherence, but this is probably hard to achieve in general, and certainly is not relevant in $Pb_{1-x}Tl_xTe$. 2) If the negative U is largely or entirely a consequence electronic correlations, then there is no reason for a large Franck-Condon suppression.

Indeed, the idea that Tl and In doping could lead to negative U centers comes from the quantum chemistry notion that these are "valence skipping elements." More or less independent of its solid state environment, in crystalline materials in which the nominal valence of Tl is +2, the symmetry between different Tl sites is always broken so that half the sites have the effective radius expected for $Tl^{+1}$ and the other half corresponding to $Tl^{+3}$. With intuition derived from an ionic picture of such solids, this is referred to as "disproportionation." This phenomenon is observed, for instance, in both In and Tl monochalgenides, as well as in AgO in which the divalent magnetic $Ag^{+2}$ disproportionates to form equal concentrations of non-magnetic $Ag^{+1}$ and $Ag^{+3}$.  Disproportionation commonly occurs when the ground state of the cation with the nominal valence would be expected to contain a half-filled s-shell.[7]  In the crystal, the disproportionation is thought [5,6] to be driven by the stability of filled shells together with the response of the polarizable lattice.

The notion that in valence skipping elements, it is largely a feature of the electronic structure, rather than the polarizability of the surrounding lattice that is responsible for the negative U is a highly non-trivial extension of the original proposal of Anderson.  The significance of this idea, and estimates of the effective U for various elements was discussed by Varma[6], who in particular proposed that the negative U associated with the valence skipping property of Bi underlies both the charge density wave formation and the mechanism of superconductivity in Pb and K doped $BaBiO_3$. In particular, Tl with a nominal +2 valence and Bi with a nominal +4 valence have the same half-filled 6S orbital, and hence share the same tendency to disproportionate to produce a mixture of sites with $6S^2$ and $6S^0$ configurations, which in the case of Bi can be thought of as a mixture of $Bi^{+3}$ and $Bi^{+5}$.  While an effective attraction which is a direct and moderately local consequence of purely repulsive microscopic interactions between electrons is somewhat counterintuitive, and its role in producing negative U centers remains controversial [8], as a point of principle the possibility of such an occurrence can be established from studies of the repulsive Hubbard model on suitable clusters; pair-binding has been shown[9] (by exact diagonalization) to occur in suitable ranges of parameters on the Hubbard square, tetrahedron (where it is particularly strong), cube, and truncated (12 site) tetrahedron.

Abstracting what is important from this history reveals that there are two essential conditions needed to turn a paired (negative U) insulator into a good superconductor:  1) The localized pair states must be resonant with the Fermi energy of the itinerant electron system.  2) The amplitude for coherent tunneling of pairs between the localized (negative U) centers and the itinerant band must be substantial. The first condition can be satisfied by tuning the chemical potential; the latter is more difficult and puts limits on the allowable strength of the electron-phonon coupling.  In Tl doped PbTe, nature is kind to us – the introduction of Tl apparently both produces the negative U centers and dopes the valence band until, for $x > x_c$, the chemical potential is such as to make the $Tl^{+1}$ and $Tl^{+3}$ states degenerate.

However, despite considerable circumstantial evidence, direct evidence of negative U centers in metallic systems is still lacking.  One possible way to obtain such direct evidence would be to use a non-superconducting layer with putative negative U centers as the normal junction between two superconductors.  If single-particle tunneling is the dominant process, the junction characteristics would be expected to satisfy the Ambegoakar-Baratoff[10] relation, $I_cR=\pi\Delta/2e$, where $\Delta$ is the superconducting gap, R is the normal state junction resistance, and $I_c$ is the critical current.  However, if the critical current reflects resonant tunneling through negative U centers, $I_cR$ can exceed this value by an arbitrarily large factor.[11] Perhaps an experiment to test this idea could be undertaken in Tl doped PbTl in structures in which the Tl concentration is modulated to define an SNS junction.

In 1987, in one of the two most highly cited papers[12] of his entire stellar career, Phil proposed that high temperature superconductivity can arise by smooth evolution with doping from a novel "spin-liquid" or "RVB" insulating phase. This idea is broadly related to the notions we have discussed up until now, in the sense that it envisages pairing to be a property of the insulating state which is inherited by the conducting state attained upon doping.[13]

There are, however, many fundamental differences between an RVB insulator and a negative U center insulator; they are, in fact, distinct phases of matter. One practical difference is that the "pairing" in the RVB state is collective, so that there is no issue of an associated Franck-Condon effect suppressing coherence as in the case of localized negative U centers. Indeed, in the original proposal of Anderson, there was no obvious energy scale associated with pairing, but in subsequent work based on the same notion[14-17], the idea emerged that in the insulating state, the pairing scale corresponds to a spin-gap or spin pseudo-gap, and that this evolves directly into the superconducting gap upon doping. In other words, the spin liquid can be thought of as a superconducting state with vanishing superfluid density,[14-19] and the pairing is more BCS-like than real-space.

The evidence that this family of ideas applies to the cuprates has been summarized in Ref. [20]. The idea that superconducting correlations can exist in an insulating phase, and that this can lead to an anomalously large pairing scale is appealing in a more general context. Since the superconducting $T_c$ is determined by the lesser of the pairing scale and the scale set by the superfluid stiffness, in any superconductor to which these ideas apply it is likely[19] that phase ordering plays an unusually important role in determining $T_c$, especially upon close approach to a putative nearly superconducting insulator. A version of such a mechanism[21] is the "spin-gap proximity effect," which is in a sense half-way between a negative U and an RVB scenario. Here, a strongly correlated insulator (of which the two-leg Hubbard ladder can be taken as the paradigmatic example), which is not necessarily in an exotic phase of matter such as a spin-liquid, but which has significant local superconducting correlations, is placed in contact with an itinerant metallic system; then, by an interaction that is formally equivalent to the proximity effect in superconducting-normal metal junctions, assuming only that the chemical potential of the two systems is such as to allow resonant tunneling of electron pairs between them, the spin-gap of the insulator gets transferred to the metal, where it becomes the superconducting gap.

A natural question is whether any of this offers guidance for new directions in the search for novel superconductors. One of us (THG) has long felt that O vacancies may serve as negative U centers in a variety of transition-metal oxides (possibly including certain cuprates), and thus might be a key ingredient in obtaining new superconductors with high superconducting transition temperatures. An example may be offered by $SrTiO_3$ [22], which is superconducting at a substantially lower concentration of carriers, 5 x $10^{17}$ cm$^{-3}$, than any other known superconductor. The Fermi energy at this low carrier density is approximately $E_F \sim 13K$, so the observed superconducting transition temperature, $T_c \sim 0.09K$, can be thought of as being moderately high, $T_c/E_F \sim 10^{-2}$. A plausible case can be made that attributes this high $T_c$ to oxygen vacancies. Notably, $SrTiO_3$ doped with other n-type dopants (such as Nb) at such low concentrations, is not superconducting. The art of manipulating O vacancies in interesting materials is in its infancy, but if it offers a route to new superconductors, it is certainly worth pursuing.

*Concerning this paper, Phil, we hope U is positive.*